\documentclass[twocolumn,trackchanges]{aastex7}
\usepackage[version=3]{mhchem}
\usepackage[subrefformat=parens]{subcaption}
\usepackage{siunitx}

\begin{document}

\title{Atmospheric Collapse and Habitability on Tidally-Locked Exoplanets}

\author[0009-0009-8470-1886]{Keigo Taniguchi}
\affiliation{Earth-Life Science Institute, Institute of Science Tokyo, Ookayama, Meguro, Tokyo 152-8550, Japan}
\affiliation{Department of Earth and Planetary Sciences, Institute of Science Tokyo, Ookayama, Meguro, Tokyo 152-8551, Japan}
\email{taniguchi@elsi.jp}

\author[0000-0001-9032-5826]{Takanori Kodama}
\affiliation{Earth-Life Science Institute, Institute of Science Tokyo, Ookayama, Meguro, Tokyo 152-8550, Japan}
\email{koda@elsi.jp}

\author[0000-0003-2260-9856]{Martin Turbet}
\affiliation{Laboratoire de M\'et\'eorologie Dynamique/IPSL, CNRS, Sorbonne Universit\'e, Ecole Normale Sup\'erieure, \\ 
PSL Research University, Ecole Polytechnique, 75005 Paris, France}
\affiliation{Laboratoire d'astrophysique de Bordeaux, Univ. Bordeaux, CNRS, B18N, allée Geoffroy Saint-Hilaire, 33615 Pessac, France}
\email{martin.turbet@lmd.ipsl.fr}

\author[0000-0003-4711-3099]{Guillaume Chaverot} 
\affiliation{Univ. Grenoble Alpes, CNRS, IPAG, 38000 Grenoble, France}
\email{guillaume.chaverot@univ-grenoble-alpes.fr}

\author[0000-0003-4808-9203]{Ehouarn Millour} 
\affiliation{Laboratoire de M\'et\'eorologie Dynamique/IPSL, CNRS, Sorbonne Universit\'e, Ecole Normale Sup\'erieure, \\ 
PSL Research University, Ecole Polytechnique, 75005 Paris, France}
\email{ehouarn.millour@lmd.ipsl.fr}

\author[0000-0001-6702-0872]{Hidenori Genda}
\affiliation{Earth-Life Science Institute, Institute of Science Tokyo, Ookayama, Meguro, Tokyo 152-8550, Japan}
\email{genda@elsi.jp}

\begin{abstract}

The habitability of terrestrial exoplanets orbiting M dwarfs is a key topic in the search for extraterrestrial life. 
The climates of these planets differ significantly from the Earth's due to their likely tidal locking, resulting in a hotter dayside and a colder nightside caused by uneven stellar irradiation.
On tidally-locked planets around the outer edge of the habitable zone (HZ), although the definition of the classical HZ requires thick \ce{CO2} atmosphere, \ce{CO2} can condense onto the surface, leading to the reduction of greenhouse effect. 
However, the dayside permanent stellar irradiation could maintain a surface liquid water area. 
The onset of atmospheric collapse and the persistence of surface liquid water are governed by global heat redistribution which is influenced by factors such as atmospheric mass, stellar irradiation, and greenhouse effects.
In this study, we used a three-dimensional global climate model to investigate the impact of atmospheric collapse on the presence of dayside surface liquid water. 
Our results indicate that surface liquid water could counter-intuitively persist despite atmospheric collapse.
This is because the loss of atmospheric \ce{CO2} weakens not only the greenhouse effect but also day-night heat transport, leading to less redistribution of the energy of dayside insolation to the nightside.
While atmospheric collapse is typically seen as an obstacle to maintaining a habitable climate, our findings suggest that it could play a positive role in sustaining surface liquid water on tidally-locked planets. 
Our work provides new light into the relationship between atmospheric collapse and planetary habitability.

\end{abstract}

\section{Introduction} \label{sec:intro}

Investigating the condition of planetary habitability is one of the target of searching for extraterrestrial lives. 
The concept of planetary habitability is based on the existence of surface liquid water on the surface because it's the best solvent. 
Also, all known life on the Earth use liquid water within their bodies to transport nutrients and energy \citep{kasting1993-01, cockell2020-01}. 
In order to maintain surface liquid water, planetary surface environments need to sustain appropriate pressure and temperature; thus investigating planetary habitability can be rephrased as examining planetary climate which can maintain surface liquid water. 

The concept of habitable zone (HZ) is defined as a region around the host star where liquid water can stably exist on the planetary surface. The HZ is quantified by the distance from the host star or stellar insolation which a planet receives \citep{kasting1993-01, kopparapu2013-02, cockell2020-01}. 
The edges of HZ have been investigated by using 1D and 3D climate models \citep[e.g.][]{kasting1993-01, abe2011-01, kopparapu2013-02, leconte2013-01, kadoya2016-01, kodama2019-01, kodama2021-01}.

The outer edge of the HZ (OHZ) where planetary surface can sustain liquid water against complete freezing is defined by \ce{CO2} maximum greenhouse limit \citep{kasting1993-01, kopparapu2013-02, von-paris2013-02}. 
\ce{CO2} is one of the greenhouse gas and common species in the atmosphere not only on the Earth. 
As partial pressure of \ce{CO2} increases, infrared absorption by \ce{CO2} increases, however Rayleigh scattering which increases planetary albedo also increases. 
Therefore, net greenhouse effect of \ce{CO2} is limited and reaching a maximum where infrared absorption and Rayleigh scattering balance. 
According to a previous study using a 1D radiative-convective equilibrium model \citep{kopparapu2013-02}, net \ce{CO2} greenhouse effect is the highest at \SI{8}{bar} of atmospheric pressure for an Earth-size planet. A planet with such an atmosphere would be completely covered by ice (snowball state) if stellar insolation is lower than $ 30 \% $ of the Solar constant which corresponds to \SI{1.7}{AU} from the present Sun. 
The OHZ has been investigated with 3D climate models by focusing on horizontal surface temperature distribution, cloud forcing, and hydrology \citep[e.g.][]{abe2011-01, wordsworth2011-01, kodama2021-01}, which are neglected effects in 1D models. In addition, the OHZ also depends on ocean circulation \citep{salazar2020-01, schmidt2022-01}, sea ice \citep{yang2020-01}, and cloud-surface equilibrium \citep{kite2021-01}.

The amount of atmospheric \ce{CO2} on the Earth is regulated by the balance between supply by volcanic degassing and removal by weathering if a negative feedback by the carbonate-silicate geochemical weathering exists \citep{walker1981-01, abbot2012-01, von-paris2013-02}. 
The carbonate-silicate weathering on the Earth acts as a negative feedback to stabilize the surface temperature and atmospheric \ce{CO2} partial pressure by decreasing the weathering rate when surface temperature decreases (and vice versa). 
Recent studies have found additional feedback which would act as positive feedback especially in case with high \ce{CO2} partial pressure \citep{graham2022-01, graham2024-01}. This feedback is associated by the weakening of water cycle which supplies the water for weathering if temperature increases.
This additional feedback on terrestrial exoplanets other than the Earth would contract the HZ under the geological time scale.

From an observational point of view, exoplanets orbiting around M dwarfs are essential targets \citep{reid1997-01, wolfgang2012-01}. 
Some of the detected exoplanets are terrestrial planets as shown by the bulk density derived from planetary mass and size \citep[e.g.][]{agol2021-01}. 
Moreover, a few of them are considered to be potentially habitable planets whose orbits are within the HZ due to their orbital distances \citep{anglada-escude2016-01, dittmann2017-01, gillon2017-01}.

Terrestrial planets around M dwarfs have different climate characteristics from the Earth's climate. 
Due to the lower luminosity of M dwarfs, the equivalent insolation is received at a smaller orbital distance than for a G star.
Consequently, these planets experience strong tidal forces, resulting in a tidally-locked state where the rotation and orbital periods synchronize \citep{barnes2017-01, bolmont2015-01, dobrovolskis2009-01, von-bloh2007-01}. 
On tidally-locked planets, planetary surface can be categorized into two hemispheres of dayside hemisphere (permanently irradiated hemisphere) and nightside hemisphere (dark, no irradiated hemisphere). 

While rapid rotators around G dwarfs like the Earth have a zonal irradiation pattern, tidally-locked planets around M dwarfs have a concentric distribution centered on the substellar point of the dayside.
This difference in irradiation leads to distinct surface temperature and liquid water distributions \citep{checlair2019-01}.
In case of cooler tidally-locked planets, strong permanent insolation persist the surface liquid water around the substellar point against the completely freezing; this kind of planet is called as eye-ball planet \citep{pierrehumbert2010-01, checlair2019-01, shields2019-01}.  
In addition, surface liquid water on tidally-locked planets is more stable than that on rapid rotators against the ice-albedo feedback which causes snowball conditions \citep{joshi2012-01, shields2013-01, checlair2017-01, checlair2019-01}. 

Besides, atmospheric circulation is different from that on the Earth \citep{ding2020-01, yang2014-01, shields2019-01}. 
Rapid rotators have equator-to-pole circulation (e.g., Hadley cells), whereas tidally-locked planets have day-night circulation.
In addition to the day-night circulation, Coriolis forces induce asymmetric circulation depending on planetary rotation rate \citep{haqq-misra2018-01}. 
Atmospheric circulation regime on tidally locked planets is classified into three types of regime according to the Rossby radius of deformation and Rhines length, which physical quantities are depending on planetary radius and rotation rate. 
These circulation differences affect climate characteristics such as cloud location, day-night heat transport, and day-night temperature contrast.

On cooler tidally-locked planets, atmospheric collapse is an important process for planetary climate, since volatile species such as \ce{CO2} begins to condense and be removed from the atmosphere \citep{wordsworth2015-01, koll2016-01, turbet2017-01, turbet2018-01, auclair-desrotour2020-01}. 
This process has been studied in the context of Mars, as it can prevent the maintenance of a strong \ce{CO2} greenhouse effect 
\citep{forget1999-01, forget2013-01, soto2015-01}. 
The onset of atmospheric collapse (\ce{CO2} condensation) depends on the surface temperature and condensation temperature of \ce{CO2}, which is a function of \ce{CO2} partial pressure. 
Previous studies investigated the condition of atmospheric collapse on tidally-locked planets using a GCM, 1-D radiative-convective model, and/or box model \citep{wordsworth2015-01, koll2016-01, turbet2017-01, turbet2018-01, auclair-desrotour2020-01}. 
The onset of atmospheric collapse, whether \ce{CO2} or volatile species condenses onto the surface or not, depends on the nightside surface temperature and condensation temperature of the volatile species, which is a function of its partial pressure. 
The nightside surface temperature is influenced by atmospheric heat transport and greenhouse effect \citep{wordsworth2015-01, koll2016-01, turbet2017-01, turbet2018-01, auclair-desrotour2020-01, ding2020-01, fan2023-01}. 
Although cooler planets near the OHZ require a substantial \ce{CO2} atmosphere for a significant greenhouse effect, atmospheric collapse can decrease \ce{CO2} greenhouse effect and surface temperature, resulting in further decreasing in atmospheric \ce{CO2} amount and surface temperature. 
Moreover, if an atmosphere with abundant \ce{CO2} experiences atmospheric collapse, the pre-collapse $p_{\ce{CO2}}$ is unlikely to recover due to the irreversible feedback of condensation. 
Consequently, the conditions at the OHZ, which necessitate a strong greenhouse effect from a massive \ce{CO2} atmosphere, may not be satisfied for tidally-locked planets. 
Atmospheric collapse would destabilize the atmosphere, which should maintain high $p_{\ce{CO2}}$ for warm environment, and alter the atmospheric condition as well as the the \ce{CO2} instability by geological cycle \citep{graham2022-01, graham2024-01}.

Atmospheric collapse on tidally-locked planets could make the HZ narrower because \ce{CO2} condensation prevents massive \ce{CO2} atmosphere which is the premise of maximum greenhouse limit, however, its influence has not been yet explored in a fully consistent 3D global climate model.
Although atmospheric collapse results in a cold climate, on tidally-locked planets, the efficiency of atmospheric heat transport could be decrease largely due to the loss of total pressure during atmospheric collapse.  
This reduction in heat transport can lead to less redistribution of strong dayside insolation to the nightside. 
To investigate the impact of atmospheric collapse on planetary habitability, we simulated climates with a 3D global climate model by changing stellar insolation.  

\section{Method} \label{sec:method}

To explore the impact of atmospheric collapse to planetary climate and habitability on tidally-locked planets, we used the Generic PCM, a three-dimensional GCM (Global Climate Model) historically developed at Laboratoire de M\'et\'eorologie Dynamique (LMD). 
The Generic PCM has previously been used to simulate solar system planet paleoclimates and climates of exoplanets \citep[e.g.][]{charnay2013-01, leconte2013-01, forget2013-01, wordsworth2015-02, turbet2017-01, turbet2018-01}. 
The Generic PCM consists of a finite difference dynamical core which solves the primitive equation of atmospheric dynamics and physical parametrizations. 
As used in \cite{turbet2018-01}, model resolution is $64 \times 48 \times 26$ in longitude $\times$ latitude $\times$ altitude with \SI{180}{s} of time step to satisfy the Courant-Friedrichs-Lewy (CFL) condition. 
In terms of physics parametrization, the time step for physics is \SI{900}{s}. 
Subgrid-scale dynamical processes such as turbulent mixing and convection is based on the parametrization produced by \cite{forget1999-01} and \cite{wordsworth2013-01}.
The planetary boundary layer is represented by the schemes by \cite{mellor1982-01} and \cite{galperin1988-01}
Rainfall and re-evaporation schemes are based on \cite{boucher1995-01}, \cite{emanuel1999-01} and \cite{leconte2013-01}. 
Land module with 18 layers calculates the thermal diffusion. These layers can represent either a rocky ground, icy ground or ocean. 
 
The radiation scheme takes into account the absorption and scattering by atmosphere, clouds, and surface in visible and infrared wavelength \citep{wordsworth2011-01}. 
The surface albedo depends on the surface condition whether it is covered by \ce{H2O} ice, \ce{CO2} ice or no ice. 
In terms of atmosphere, this scheme treats \ce{CO2} and \ce{H2O} absorption using correlated-$k$ method with two-streams approximation for both visible and infrared range of wavelength \citep{fu1992-01, toon1989-01}. 

\ce{CO2} condensation scheme is included in our calculations. 
This scheme calculates \ce{CO2} condensation and sublimation in the atmosphere and on the surface using \ce{CO2} partial pressure and temperature \citep{fanale1982-01, wordsworth2015-01}. 
The formula of \ce{CO2} condensation temperature $T_{\mathrm{cond}, \ce{CO2}} \,[\si{K}]$ as a function of \ce{CO2} partial pressure $p_{\ce{CO2}} \,[\si{Pa}]$ for less than the \ce{CO2} triple-point value ($p_{\ce{CO2}} \le 518000 \,\si{Pa}$) 
\begin{equation}
    T_{\mathrm{cond}, \ce{CO2}} = \frac{-3167.8}{\ln{[0.01\, p_{\ce{CO2}}]} - 23.23}. 
\end{equation}
The $p_{\ce{CO2}}$ in our all simulations is lower than the pressure at the \ce{CO2} triple-point. 
According to this equation, the removal of atmospheric \ce{CO2} and changing in temperature by \ce{CO2} latent heat are calculated. 
Radiative effect of \ce{CO2} cloud is not included as in \cite{turbet2018-01} because of low radiative effect due to cooler low mass host star and partial cloud coverage \citep{forget1999-01, forget2013-01, kitzmann2017-01}.
The surface emissivity is depending on whether the surface is covered by \ce{H2O} ice, \ce{CO2} ice, or no ice. 
If a grid is completely covered by \ce{H2O} ice, its surface emissivity is 1.0, and if covered by \ce{CO2} ice, it is overwritten to 0.9 of \ce{CO2} ice's value as in \cite{turbet2017-01}.


In this study, we assumed a tidally-locked Earth-size planet with circular orbit (obliquity and eccentricity are zero) to remove the seasonal changes in stellar insolation. 
In addition, we assume an aqua planet, which type of planet is covered by \ce{H2O} ocean. 
The thermal inertia in the land module, which represents the resistance of ground to temperature change, is fixed to $ 20000 $ \si{J m^{-2} K^{-1} s^{-0.5}} to mimic the effect of a well-mixed surface ocean layer \citep{cheruy2017-01, turbet2018-01}, but oceanic heat transport are not included. 

In the initial state, the planetary surface is covered by 1 m of \ce{H2O} ice with \SI{210}{K} of temperature. 
The surface water (both ice and liquid) exchanges to atmosphere, but doesn't flow from a grid to its neighbors. 
The surface \ce{H2O} ice in the initial condition is expected to eventually melt on the dayside if the dayside is enough heated. 
In addition, the distribution of surface \ce{H2O} ice (and liquid \ce{H2O}) when the GCM reaches the stable state would be less affected by the initial surface condition because of the less importance for ice-albedo feedback around M dwarfs \citep[e.g.][]{checlair2017-01}. 
We assume static, isothermal atmosphere with \SI{210}{K} and no winds as initial condition of the atmosphere. 

Input parameters for a series of simulations are stellar insolation $S_{p}$ (the percentage compared to the Solar constant $S_{0} = \SI{1366}{Wm^{-2}}$), \ce{N2} partial pressure $ p_{\ce{N2}}$, and \ce{CO2} partial pressure $ p_{\ce{CO2}}$, summarized in Table \ref{tab:inputs}. 
To simulate a climate on tidally-locked planet orbiting a M dwarf, we set a TRAPPIST-1 as the host star. 
We set host star's spectrum data and planet's rotation period from TRAPPIST-1's observational data. 
The relationship between stellar insolation and rotation rate $\Omega$ (corresponding to the orbital period) was derived from 
\begin{equation}
    \Omega = C S_{\mathrm{p}}^\frac{3}{4}
\end{equation}
using Kepler's laws. 
The coefficient $C = \num{7.385e-8}$ is derived from the observational data \citep{agol2021-01}. 
The rotation rates $\Omega$ in case of $0.3, 0.4, 0.5 \, S_{0}$ are $ 6.72, 8.34, 9.86 \times 10^{-6}$ \si{s^{-1}}, respectively. 
Our parameter range of stellar insolation is similar to TRAPPIST-1f which receives $0.37 \, S_{0}$ orbiting in a period of $9.2$ Earth days. 

\begin{table}
    \centering
    \footnotesize
    \begin{tabular}{lcc} \hline
        $ S_{\mathrm{p}} $   & $ 0.3 \, S_{0}, 0.4 \, S_{0}, 0.5 \, S_{0} $ & $ S_{0} = $ \SI{1366}{W m^{-2}} \\ 
        $ \Omega $       &  6.72, 8.34, 9.86        & $ \times 10^{-6} $ \si{s^{-1}} \\
        $ P $       &  10.8, 8.71, 7.37        & \si{days} \\
        $ p_{\ce{N2}}$  & $ 0.1, 1.0 $ & \si{bar} \\
        $ p_{\ce{CO2}}$ & $ 10^{-6}, 10^{-5}, 10^{-4}, 10^{-3}, 10^{-2}, 0.1, 1.0 $ & \si{bar} \\ \hline
    \end{tabular}
    \caption{Input parameters for the simulation}
    \label{tab:inputs}
\end{table}
\setcounter{footnote}{0}
Regarding to the $ p_{\ce{N2}}$ and $ p_{\ce{CO2}}$, we set correlated-$k$ tables for each composition using SpeCT and Exo\_k. 
SpeCT (SPEctra for correlated-$k$ Tables) \citep{chaverot-2023} calculates the hundreds of spectra required to produce a correlated-$k$ table by using HITRAN 2020\footnote{\url{https://hitran.org/}}  \citep{rothman2005-01,gordon2022-01}. 
Experimental correction factors are included in SpeCT to calculate accurately the \ce{CO2} line profiles, as well as the continua \citep{chaverot-2023}.
We use the MT\_CKD database for water continua \citep{mlawer_development_2012}.

Exo\_k\footnote{\url{https://perso.astrophy.u-bordeaux.fr/~jleconte/exo_k-doc/index.html}} is an open-source python library to compute correlated-$k$ coefficient of compound atmosphere \citep{leconte2021-01}. 
We created correlated-$k$ tables for Generic PCM with Exo\_k inputting high-resolution spectra of \ce{CO2} and \ce{H2O} from SpeCT, and collision-induced absorptions (CIA) and dimer absorption \citep{gruszka1997-01, baranov2004-01, wordsworth2010-01, richard2012-01}. 
The CIA tables we included are $\ce{N2}\mathchar`-\ce{N2}$, $\ce{CO2}\mathchar`-\ce{CO2}$, $\ce{H2O}\mathchar`-\ce{H2O}$, $\ce{H2O}\mathchar`-\ce{CO2}$, and $\ce{H2O}\mathchar`-\ce{N2}$. 
We set the resolution of pressure grid ($ P = \{1.0, 10^{1}, 10^{2}, 10^{3}, 10^{4}, 10^{5}, 10^{6}, 10^{7}  \, \si{Pa} \} $), temperature grid ($ T = \{50, 110, 170, 230, 290, 350 \, \si{K} \} $, and \ce{H2O} mixing ratio grid ($ Q = \{10^{-7}, 10^{-6}, 10^{-5}, 10^{-4}, 10^{-3}, 10^{-2}, 10^{-1}, 1.0\} $).

We run the GCM for each case of stellar insolation $S_{\mathrm{p}}$ (and corresponding $ \Omega $), $ p_{\ce{N2}}$, and $ p_{\ce{CO2}}$ following the time variation of temperature and radiative balance. 
When surface temperatures are stable and both incoming stellar flux and outgoing radiation balance within $2 \text{\textendash} 3$ \si{W m^{-2}}, we regard that the climate is in a quasi-equilibrium state. 
The duration until the climate becomes in a stable state is $10 \text{\textendash} 20$ Earth's years and the last 1-year averaged data is used for analysis.
The onset of atmospheric collapse is determined by the amount of \ce{CO2} ice on the surface. 
If the \ce{CO2} ice on the surface is zero everywhere and the minimum surface temperature is higher than \ce{CO2} sublimation temperature, we consider that atmospheric collapse is not occurring in the case. On the other hand, if the amount of \ce{CO2} ice on the surface increases over time, atmospheric collapse is occurring. 
Note that if \ce{CO2} condensation occurs and the condensation speed is high enough that the decreasing in $ p_{\ce{CO2}}$ changes the condensation temperature (equal to the minimum surface temperature) during the simulation, we didn’t select the last 1-year data at the end of the calculation but selected 1-year data from the beginning of the onset of atmospheric collapse. 
This is because in the model setup we used, the GCM can treat at most only one species (\ce{H2O} in these calculations) as a variable gas in order to reflect the change of atmospheric composition to the moist convection and radiative transfer; thus, the GCM cannot correctly reflect the rapid change of \ce{CO2} mixing ratio during the atmospheric collapse. 

\section{Results} \label{sec:result}

In this section, we firstly show typical climatic behavior, such as surface temperature and circulation pattern in case of where both atmospheric collapse occurs and does not. 
Then, we summarize all the cases and represent the fate of climates and habitability where atmospheric collapse occurs by connecting the cases.

\subsection{Onset of atmospheric collapse}

\begin{figure*}[htbp!]
\begin{center}
\includegraphics[scale=0.90]{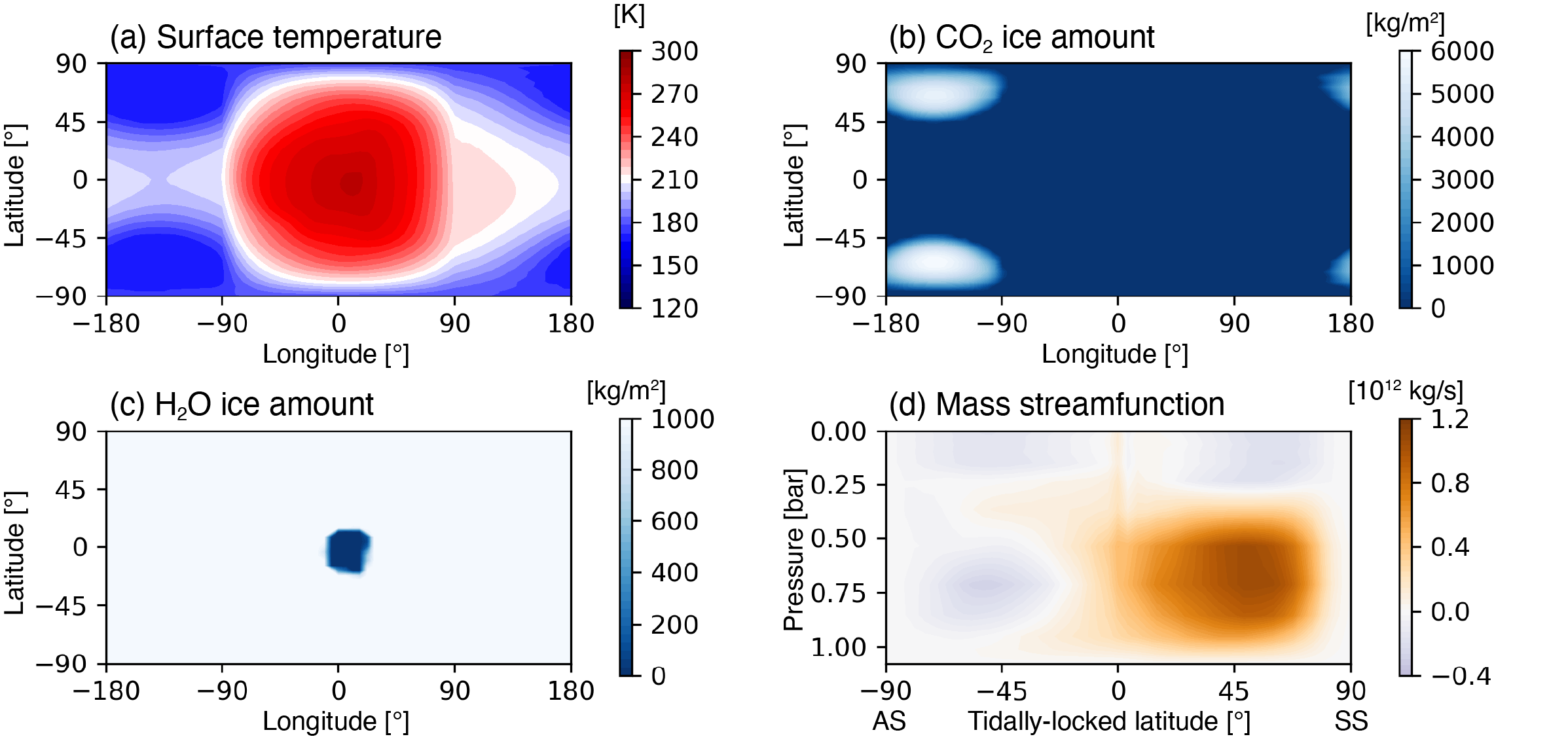}
\caption{Atmospheric conditions in the case with $S_{\mathrm{p}} = 0.4 \, S_{0}$, $p_{\ce{N2}} = \SI{1.0}{bar}$, $p_{\ce{CO2}} = \SI{0.1}{bar}$. Each case shows (a) horizontal distribution of surface temperature, (b) horizontal distribution of \ce{CO2} ice amount (c) horizontal distribution of \ce{H2O} ice amount, and (d) mass streamfunction between the substellar point (SS) and antistellar point (AS) in tidally-locekd coordinate. In (a), (b), and (c), 0\textdegree of latitude and of longitude corresponds to the substellar point (SS). Note that horizontal axis in (d) represents the latitude in tidally-locked coordinate and the mass flux where the value is positive is anticlockwise circulation.}
\label{fig:tv_C1e-01}
\end{center}
\end{figure*}

Figure \ref{fig:tv_C1e-01} shows the results in a case with $S_{\mathrm{p}} = 0.4 \, S_{0}$ (rotation rates $\Omega$ and rotation period $P$ are $ 8.31 \times 10^{-6} $ \si{s^{-1}} and $8.71$ \si{days}, respectively), $p_{\ce{N2}} = \SI{1.0}{bar}$, $p_{\ce{CO2}} = \SI{0.1}{bar}$: (a), (b), and (c) are the horizontal distribution of surface temperature, surface \ce{CO2} ice amount, and surface \ce{H2O} ice amount, respectively, and (d) shows the mass streamfunction between the substellar point (SS) and antistellar point (AS) in tidally-locked coordinate \citep{koll2015-01}. 
Note that the vertical orange line at 0\textdegree of tidally-locked latitude would be a numerical issue because the area contains the north/south poles, which corresponds to coordinate singularity in the GCM. 

The surface temperature around the SS point is higher than \SI{273}{K}, resulting in forming partial habitable area where surface liquid water exists (Figure \ref{fig:tv_C1e-01} (c)). 
Note that the melting point of \ce{H2O} does not depend much on the pressure in contrast to the condensation temperature of \ce{CO2}. 
Besides, strong convection induced by the permanent stellar insolation makes the day-night circulation (Figure \ref{fig:tv_C1e-01} (d)).
The temperature distribution and direct day-night circulation are consistent with slow rotator regime described in \cite{haqq-misra2018-01}. 
On the other hand, the minimum surface temperature on the nightside doesn't locate at the antistellar point, but in the higher latitudes because of the lower latitudes heating by equatorial flow and formation of nightside gyre in the higher latitudes which prevents the supply of heat into the gyre \citep{showman2010-01}. 
In this case, the surface temperature in the higher latitudes is low enough that atmospheric \ce{CO2} can condense, and then the temperature is fixed to \SI{170}{K} which is equal to the \ce{CO2} condensation temperature of \SI{0.1}{bar} of \ce{CO2} \citep{fray2009-01}. 
As a result, surface \ce{CO2} ice distributes the higher latitudes on the nightside (Figure \ref{fig:tv_C1e-01} (b)), and atmospheric \ce{CO2} is expected to decrease. 

\begin{figure*}[htbp!]
\begin{center}
\includegraphics[scale=0.90]{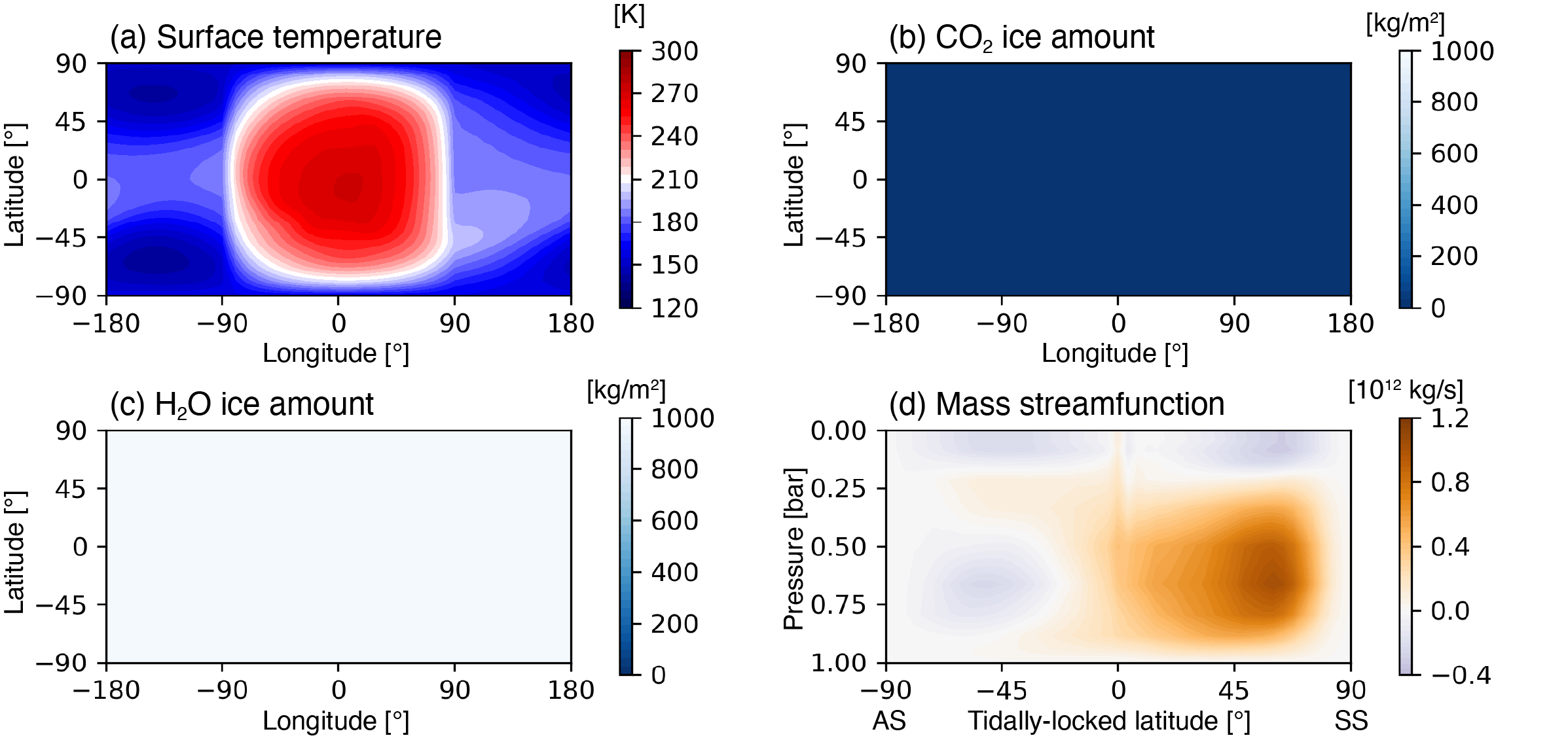}
\caption{Atmospheric conditions in the case with $S_{\mathrm{p}} = 0.4 \, S_{0}$, $p_{\ce{N2}} = \SI{1.0}{bar}$, $p_{\ce{CO2}} = 10^{-4} \, \si{bar}$ as in Figure \ref{fig:tv_C1e-01}; (a) horizontal distribution of surface temperature, (b) horizontal distribution of \ce{CO2} ice, (c) horizontal distribution of \ce{H2O} ice, and (d) mass streamfunction between the substellar point and antistellar point.}
\label{fig:tv_C1e-04}
\end{center}
\end{figure*}

Figure \ref{fig:tv_C1e-04} shows the results in a case with $S_{\mathrm{p}} = 0.4 \, S_{0}$, $p_{\ce{N2}} = \SI{1.0}{bar}$, $p_{\ce{CO2}} = 10^{-4} \, \mathrm{bar}$ as shown in Figure \ref{fig:tv_C1e-01}. 
The coldest point locates in the higher latitudes in the nightside as in a case with $p_{\ce{CO2}} = \SI{0.1}{bar}$, and the minimum surface temperature is lower due to the lower \ce{CO2} greenhouse effect.  
However, the minimum surface temperature is higher than \SI{124}{K} which is the condensation temperature of $p_{\ce{CO2}} = 10^{-4} \, \mathrm{bar}$, and thus there is no surface \ce{CO2} ice and atmospheric collapse in this case. 
On the other hand, the maximum surface temperature is lower than \SI{273}{K}, resulting in completely freezing surface (snowball state).

\subsection{The fate of \ce{CO2} partial pressure}

\begin{figure*}[htbp!]
\begin{center}
\includegraphics[scale=0.80]{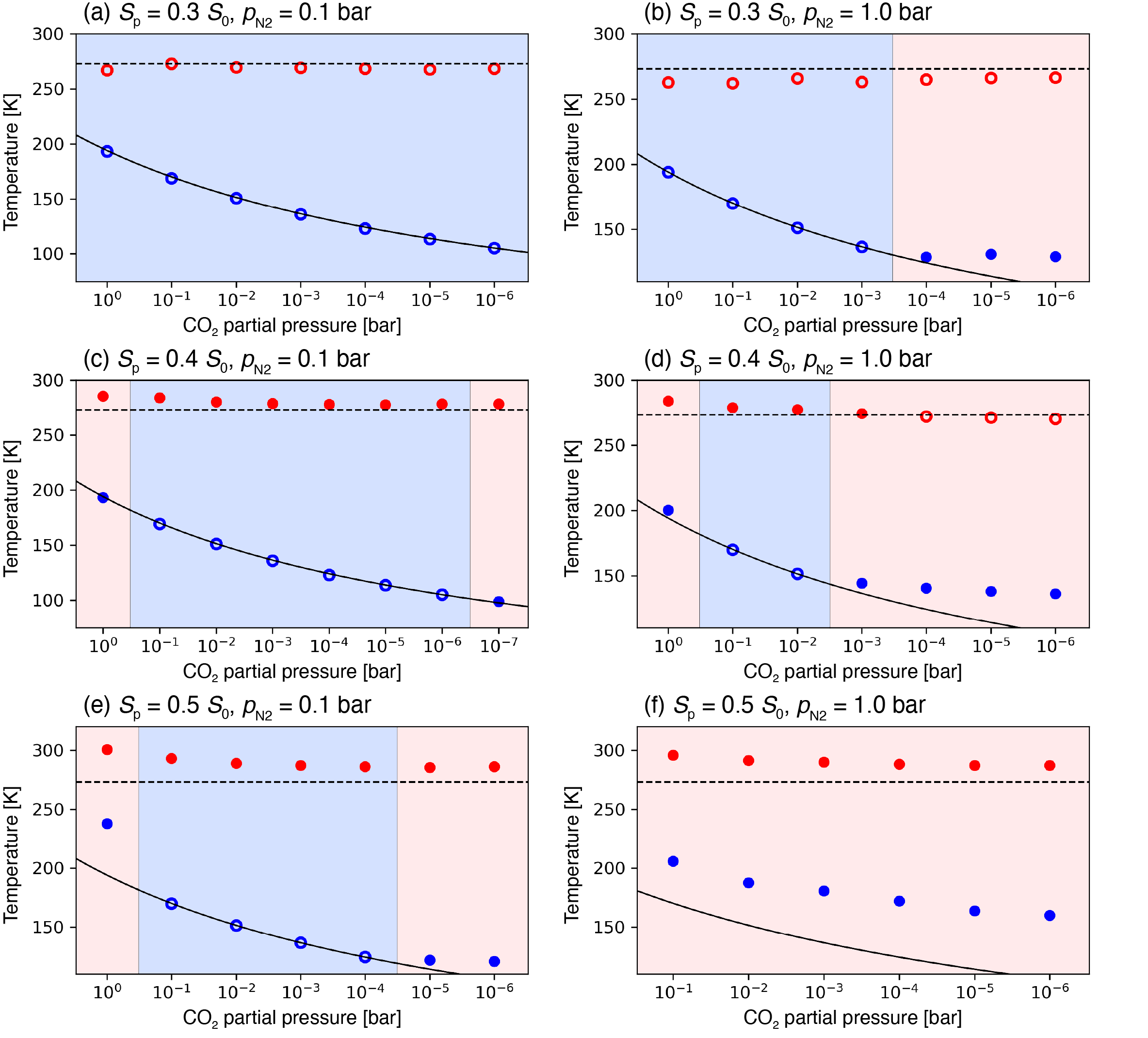}
\caption{Maximum (red dots) and minimum (blue dots) surface temperatures and the onset of atmospheric collapse of each case. The red dots shows that the maximum surface temperature is higher than the melting point of \ce{H2O} (\SI{273}{K}), and on the other hand, red circles show that the surface is completely freezing. The black curve indicates the condensation temperature of \ce{CO2} as a function of $p_{\ce{CO2}}$. The blue dots indicate that the minimum surface temperature is higher than the condensation temperature (black curve). The blue circles show that the atmospheric collapse is occurring and thus the circles locate on the black curve of condensation temperature. The red/blue shaded areas are corresponding to the cases where minimum surface temperature is above/equal to \ce{CO2} condensation temperature. }
\label{fig:collapse}
\end{center}    
\end{figure*}

As in the previous section, we examined the onset of atmospheric collapse and habitability by checking the surface temperatures for each case (listed in Table \ref{tab:inputs}). 
Figure \ref{fig:collapse} shows the minimum and maximum surface temperatures of the cases with (a) $S_{\mathrm{p}} = 0.3 \, S_{0}$, $p_{\ce{N2}} = \SI{0.1}{bar}$, (b) $S_{\mathrm{p}} = 0.3 \, S_{0}$, $p_{\ce{N2}} = \SI{1.0}{bar}$, (c) $S_{\mathrm{p}} = 0.4 \, S_{0}$, $p_{\ce{N2}} = \SI{0.1}{bar}$, (d) $S_{\mathrm{p}} = 0.4 \, S_{0}$, $p_{\ce{N2}} = \SI{1.0}{bar}$, (e) $S_{\mathrm{p}} = 0.5 \, S_{0}$, $p_{\ce{N2}} = \SI{0.1}{bar}$, and (f) $S_{\mathrm{p}} = 0.5 \, S_{0}$, $p_{\ce{N2}} = \SI{1.0}{bar}$. 
The results of Figure \ref{fig:tv_C1e-01} and Figure \ref{fig:tv_C1e-04} are corresponding to the two points of maximum and minimum surface temperatures in Figure \ref{fig:collapse} (d).

In terms of minimum surface temperatures in Figure \ref{fig:collapse}, the blue dots/circles represent the cases where the minimum surface temperatures are above/equal to \ce{CO2} condensation temperature (shown as black curves), respectively. 
Since atmospheric \ce{CO2} condenses onto the surface when surface temperature is \ce{CO2} condensation temperature, the blue dots and circles represent no condensation and occurrence of atmospheric collapse, respectively. 
The blue shaded areas show $p_{\ce{CO2}}$ range where atmospheric collapse occurs. 
Regarding to the case of $p_{\ce{CO2}} = \SI{0.1}{bar}$ in Figure \ref{fig:collapse} (d), atmospheric collapse occurs as shown in Figure \ref{fig:tv_C1e-01}.
If atmospheric \ce{CO2} decreases quasi-statically, $p_{\ce{CO2}}$ decrease to $ 10^{-2} $ \si{bar} after a certain length of time. 
In addition, the case with $ 10^{-2} $ \si{bar} also show that atmospheric collapse occurs. 
Thus, $p_{\ce{CO2}}$ would decrease until atmospheric collapse stops and the endpoint locates between $ 10^{-2} $ and $ 10^{-3} $ \si{bar} because of no \ce{CO2} condensation in case with $ 10^{-3} $ \si{bar}. 
To summarize, if atmospheric collapse occurs, $p_{\ce{CO2}}$ would be between $ 10^{-2} $ and $ 10^{-3} $ \si{bar} after the atmospheric collapse event. 
In the other cases where $p_{\ce{CO2}} $ is higher than \SI{0.1}{bar} or lower than $ 10^{-3} $ \si{bar}, minimum surface temperatures are higher than the condensation temperature and \ce{CO2} condensation does not occur.  
Thus, an atmosphere which initially contains $p_{\ce{CO2}}$ more than \SI{0.1}{bar} or lower than $ 10^{-3} $ \si{bar} remains the initial $p_{\ce{CO2}}$. 

In terms of maximum surface temperatures on the dayside, the red dots and circles represent the cases where the temperatures are above/below the \ce{H2O} melting point (shown as dashed lines), respectively. 
Besides, in all cases with higher maximum surface temperature than melting point, surface liquid water exists around the substellar point, but no surface liquid water in the other cases with lower maximum surface temperature than melting point. 
Interestingly, the maximum surface temperatures in Figure \ref{fig:collapse} (d) are higher than the melting point even in cases where atmospheric collapse is occurring.
Namely, it means that locally habitable environment on the dayside remains during/after atmospheric collapse against the decreasing in \ce{CO2} greenhouse effect. 
This scenario can be seen in case of $S_{\mathrm{p}} = 0.4 \, S_{0}$, $p_{\ce{N2}} = \SI{0.1}{bar}$ in Figure \ref{fig:collapse} (c) despite the $p_{\ce{CO2}}$ range of atmospheric collapse is different. 

In the cases with $S_{\mathrm{p}} = 0.3 \, S_{0}$ (Figure \ref{fig:collapse} (a) and (b)), the surface is globally covered with ice regardless of the  $p_{\ce{CO2}}$ and the onset of atmospheric collapse. 
On the other hand, in the cases with $S_{\mathrm{p}} = 0.5 \, S_{0}$ (Figure \ref{fig:collapse} (e) and (f)), surface liquid water in all cases exists around the substellar point.

\section{Discussion} \label{sec:discussion}

\subsection{Heat redistribution and day-night surface temperatures}


\begin{figure*}[htbp!]
\begin{center}
\includegraphics[scale=0.80]{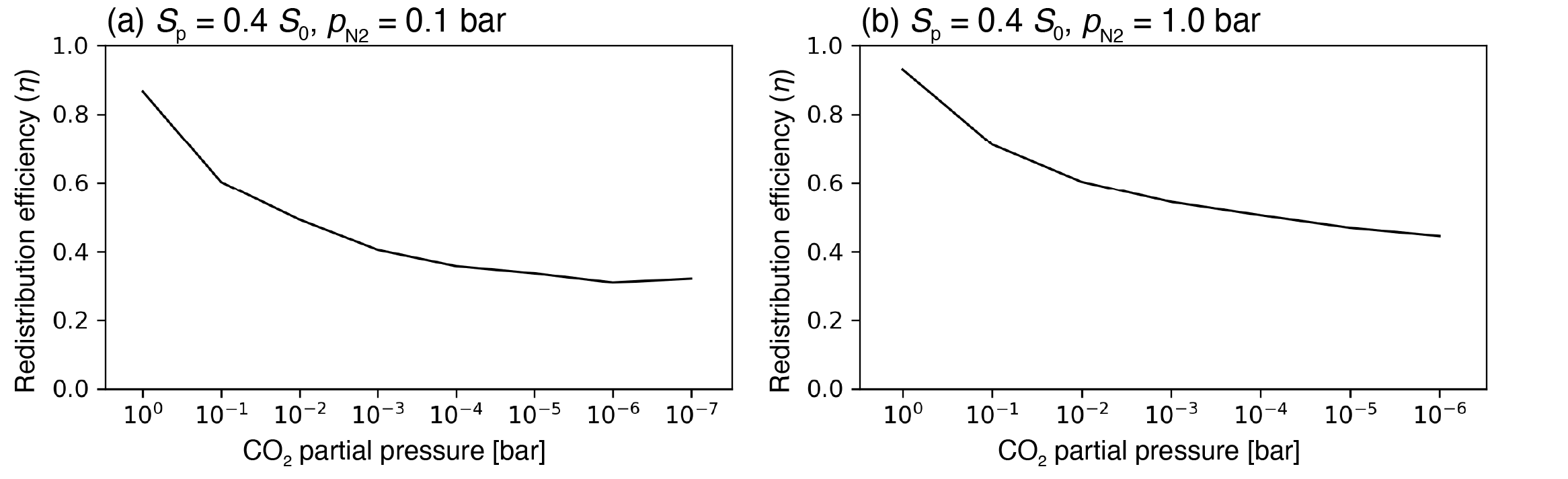}
\caption{Thermal redistribution efficiency calculated by the day/nightside outgoing longwave radiation (OLR) $ \eta = OLR_{\mathrm{nightside}}/OLR_{\mathrm{dayside}} $ in case of (a) $S_{\mathrm{p}} = 0.4 \, S_{0}$, , $p_{\ce{N2}} = \SI{0.1}{bar}$, (b) $S_{\mathrm{p}} = 0.4 \, S_{0}$, , $p_{\ce{N2}} = \SI{1.0}{bar}$. }
\label{fig:redistribution}
\end{center}    
\end{figure*}

Surface temperature distribution on tidally-locked planets is mainly controlled by insolation distribution and global heat transport from dayside to nightside. 
Previous studies investigated the day-night heat redistribution and atmospheric collapse under \ce{CO2}-dominant atmosphere by using analytical models, which predict the surface temperature and heat transport by solving the heat flux such as radiation and sensible heat transport on dayside surface, nightside surface, and atmosphere  \citep{wordsworth2015-01, auclair-desrotour2020-01}. 
They proved that the nightside surface temperature, which determines the condition of atmospheric collapse can be attributed to radiative effects (longwave/shortwave absorption) and dynamical effects (such as horizontal heat transport and sensible heat). 

As mentioned in the previous section, dayside surface temperature remains during the atmospheric collapse while decreasing the $p_{\ce{CO2}}$. 
The maintenance mechanism of dayside surface temperature is related to the changing in the heat transport during decreasing the $p_{\ce{CO2}}$.  
To investigate the day-night, horizontal heat transport in our GCM simulations, we calculated the thermal redistribution efficiency $\eta$, which is defined as the ratio of outgoing longwave radiation (OLR) on the nightside to that on the dayside \citep{leconte2013-01}. 
If $ \eta  = 1.0 $, absorbed insolation on the dayside is globally distributed efficiently, and if $ \eta = 0 $, the atmosphere doesn't transport any energy to the nightside. 
Figure \ref{fig:redistribution} shows the thermal redistribution efficiency as a function of $p_{\ce{CO2}}$ in case with (a) $S_{\mathrm{p}} = 0.4 \, S_{0}$, $p_{\ce{N2}} = \SI{0.1}{bar}$, (b) $S_{\mathrm{p}} = 0.4 \, S_{0}$, $p_{\ce{N2}} = \SI{1.0}{bar}$. 
In both $p_{\ce{N2}}$ cases, the thermal redistribution efficiency is high $\eta \approx 0.9$ around $1.0 \si{bar}$ of $p_{\ce{CO2}}$, and $\eta \approx 0.4$ around $10^{-6} \si{bar}$ of $p_{\ce{CO2}}$. 
The change in thermal redistribution efficiency seems to be the shift from nightside atmospheric cooling at high $p_{\ce{CO2}}$ to nightside surface cooling at low $p_{\ce{CO2}}$ rather than the change in dynamical effect by reducing the total pressure of atmosphere \citep{auclair-desrotour2020-01, wang2022-01}. 
If the change in total pressure played a dominant role, the change in $\eta$ would be sensitive to the total mass of the atmosphere. 
The changing in $\eta$ would be a lot in case of $p_{\ce{N2}} = \SI{0.1}{bar}$ (the range of total pressure $p_{\mathrm{s}}$ is between \SI{1.1}{bar} and \SI{0.1}{bar}) and  the changing in $\eta$ would be low $p_{\ce{N2}} = \SI{1.0}{bar}$ (the range of total pressure $p_{\mathrm{s}}$ is between \SI{2.0}{bar} and \SI{1.0}{bar}). 

Weakened atmospheric heat redistribution as decreasing $p_{\ce{CO2}}$ keeps the insolation energy on the dayside, which should be transported to the nightside. 
Besides, the weakened atmospheric heat transport makes day-night temperature contrast larger. 
Maintaining the dayside surface temperature above $273 \si{K}$ even during/after atmospheric collapse while nightside surface temperature decreases is corresponding to the behavior of atmospheric heat redistribution. 

While changing in total pressure seems to be less important for the thermal redistribution efficiency $\eta$, other dynamical effects such as the heat transport driven by the temperature gradient could affect to $\eta$. To quantitatively assess the shift in radiative/dynamical effect, atmospheric heat redistribution, and surface temperature, using a simplified analytical model is beneficial.
Previous studies in \cite{wordsworth2015-01, auclair-desrotour2020-01} provide the quantitative comparison and numerical predictions, however, these analytical models, which calculate the radiative heat flux and sensible heat transport, could not be directly applied to the atmosphere in our study because the atmosphere we focus consists of not only \ce{CO2} but also \ce{N2}. 
Atmospheric \ce{N2} has influences to the sensible heat transport, but does not affect to the atmospheric radiative property.
In addition to the $p_{\ce{N2}}$ dependency, heat transport between the nightside surface and atmosphere, which is corresponding to the inversion layer, should be carefully considered. Heat transport and mixing process in inversion layer remain a complex subject even on the Earth \citep{joshi2020-01}. 
Since it is needed to include the $p_{\ce{N2}}$ dependency and heat transport scheme in inversion layer in an analytical model, we will develop a new analytical model for \ce{N2} and \ce{CO2} atmosphere, and will do numerical predictions of radiative flux, sensible heat flux, and dayside/nightside surface temperature as a future work. 

\subsection{Surface liquid water and water cycle}

The amount of surface water on exoplanets around M dwarfs derived during the planetary formation is thought to have wide range  \citep{tian2015-02, kimura2022-01}. 
One type of planets with limited surface water is known as dry planets, and the other type with large amount of water tens or hundreds times larger than the Earth ocean mass is known as ocean planets \citep{leger2004-01}. 
In this study, we set a slab ocean on the planetary surface, which acts as a \ce{H2O} storage but doesn't have a role of heat transport. The condition of surface water simulated in GCM would be different from the actual surface water on dry/ocean planets. In addition,  we showed that the surface temperature around the substellar point is higher than the \ce{H2O} melting point of $273 \si{K}$ during/after atmospheric collapse, but it doesn't indicate the existence of water there on an actual planetary surface. We discuss the behavior of surface water on dry/ocean planets other than the slab ocean in this section. 

For dry planets, \cite{ding2020-01} investigated the surface water redistribution on tidally-locked planets using a GCM, and they found that the surface water redistribution depends on the specific humidity at substellar tropopause $q_{\mathrm{sat}, \mathrm{d}}$, and nightside surface $q_{\mathrm{sat}, \mathrm{n}}$. In case with low $p_{\ce{CO2}}$ (\ce{N2} atmosphere and $400 \si{ppm}$ of \ce{CO2}), $q_{\mathrm{sat}, \mathrm{n}} < q_{\mathrm{sat}, \mathrm{d}}$ , resulting in the water redistribution from dayside to the nightside. As a result, surface water exists as nightside ice while dayside surface being dry. On the other hand, in case with high $p_{\ce{CO2}}$ (with $0.1 \si{bar}$ of \ce{CO2}), $q_{\mathrm{sat}, \mathrm{d}} < q_{\mathrm{sat}, \mathrm{n}}$, namely, dayside surface water evaporated is cold-trapped at the substellar tropopause. The dayside surface liquid water stably exists in this condition. 

Regarding our GCM results, the behavior of $q_{\mathrm{sat}, \mathrm{d}}$ and $q_{\mathrm{sat}, \mathrm{n}}$ under high/low $p_{\ce{CO2}}$ is consistent with the conclusion in \cite{ding2020-01}: $q_{\mathrm{sat}, \mathrm{n}} < q_{\mathrm{sat}, \mathrm{d}}$ in case with $p_{\ce{CO2}} < 0.01 \ce{bar}$, and $q_{\mathrm{sat}, \mathrm{d}} < q_{\mathrm{sat}, \mathrm{n}}$ in case with $p_{\ce{CO2}} = 0.1 \ce{bar}$ and $1.0 \ce{bar}$. If the amount of surface liquid water is limited (corresponding to dry planets), surface liquid water can stably exists in case with $p_{\ce{CO2}} = 0.1 \ce{bar}$ and $1.0 \ce{bar}$, while surface water will be redistributed to the nightside as ice in case with $p_{\ce{CO2}} < 0.01 \ce{bar}$.

Surface water ice trapped at nightside in case with low $p_{\ce{CO2}}$ may not remain trapped there permanently. Geothermal heat causes to the basal melting beneath the nightside ice, leading to the ice flow from nightside to dayside \citep{yang2014-02}.
Whether surface liquid water exists on dayside in case low $p_{\ce{CO2}}$ would depend on the ice flow induced by geothermal heat.

For ocean planets, ocean heat transport can change the condition where surface temperature exceeds the melting point. 
Ocean can transport heat as well as the atmosphere, resulting in increasing the thermal redistribution efficiency and decreasing the day-night temperature contrast. 
However, ocean heat transport depends on such as disturbance by land distribution \citep{salazar2020-01, schmidt2022-01} and sea ice drift \citep{yang2020-01}. 
In addition to the heat transport, ocean as a \ce{CO2} storage makes the climatic behavior more complex. Additional processes of \ce{CO2} cycle (i.e. atmosphere-ocean \ce{CO2} exchange, burying the \ce{CO2} ice into the ocean, and formation of \ce{CO2} ice clathrate) should be considered. Since the impact of ocean circulation and whether open-sea appears on the surface is complex, simulating the dynamic ocean by GCMs is required for understanding the heat redistribution on ocean-covered tidally-locked planets.

\subsection{$p_{\ce{N2}}$ dependency on the onset of atmospheric collapse}

In this study, we set $0.1 \si{bar}$ and $1.0 \si{bar}$ of \ce{N2} as a non-condensable species, which is a similar value of $p_{\ce{N2}}$ in Earth's atmosphere. 
Although we set $p_{\ce{N2}}$,  terrestrial exoplanets can retain other values of $p_{\ce{N2}}$. 
By comparing Figure \ref{fig:collapse} (c) to (d), if $p_{\ce{N2}}$ is more lower (i.e. $p_{\ce{N2}} \le 0.01 \si{bar}$ or no background atmosphere), the blue shaded area where atmospheric collapse occurs expands and then the atmospheric \ce{CO2} will converge to full collapse. 
The lower $p_{\ce{N2}}$ atmosphere would have larger day-night temperature contrast if $p_{\ce{CO2}}$ is low due to the atmospheric collapse, resulting in colder nightside corresponding to the lower \ce{CO2} condensation pressure. 
In this scenario, the planet would resemble to an air-less planet which lost \ce{CO2} dominant atmosphere as suggested in \cite{wordsworth2015-01}. 
On the other hand, in case with higher $p_{\ce{N2}}$ (i.e. $p_{\ce{N2}} \ge 10 \si{bar}$), the blue shaded area in Figure \ref{fig:collapse} where \ce{CO2} condenses would disappear. 
Opposite from the low $p_{\ce{N2}}$ atmosphere, higher $p_{\ce{N2}}$ would contribute to the lower day-night temperature contrast, resulting in warmer nightside temperature which prevents the onset of atmospheric collapse. 
In addition, the effect of Rayleigh scattering under higher $p_{\ce{N2}}$ atmosphere would be more important in general, however, it would be less important for exoplanets around a cool star due to the because the spectrum of stellar insolation is longer \citep{paradise2021-01}. 
Higher $p_{\ce{N2}}$ atmosphere could remain atmospheric \ce{CO2} and habitable environment without occurring atmospheric collapse event.
In order for predicting the scenario with low $p_{\ce{N2}}$, we tried to do the GCM simulations with $p_{\ce{N2}} = 0.01 \si{bar}$ as $p_{\ce{N2}} = 0.1 \si{bar}$ and $1 \si{bar}$, but could not obtain the results because the some of simulations were crashed. 
Rapidly decreasing in $p_{\ce{CO2}}$ and total mass of atmosphere due to the \ce{CO2} condensation seems to prevent the simulation avoiding the computational instability. 
In addition to the rapid condensation, there is a limitation of correlated-$k$ table, which stores radiative property of the atmosphere. The correlated-$k$ table we used in this study can reflect the change in only \ce{H2O} mixing ratio. Since change in both \ce{H2O} and \ce{CO2} mixing ratio should be considered when atmospheric \ce{CO2} rapidly decreases, it is necessary to develop a new correlated-$k$ table for multi-species condensation.

\subsection{Maintenance mechanisms of surface liquid water on planets around other M dwarfs}

In this study, we assumed tidally-locked planets around TRAPPIST-1, and then found that surface liquid water can remain if atmospheric collapse occurs.
This maintenance scenario, which requires the onset of atmospheric collapse and the existence of surface liquid water, depends on surface temperature distribution especially nightside minimum surface temperature and dayside maximum surface temperature.

According to \cite{haqq-misra2018-01}, circulation regime influences atmospheric behavior such as wind and temperature distribution. 
Circulation regimes are classified by the Rossby deformation radius $\lambda_{\mathrm{R}}$ and Rhines length $L_{\mathrm{R}}$ (both depend on rotation period) compared to the planetary radius $R$. 
The ratio of the Rossby deformation radius and Rhines length to the planetary radius in our simulations were about $1.1 \le \lambda_{\mathrm{R}}/R \le 1.2$ and $1.1 \le L_{\mathrm{R}}/R \le 2.0$, respectively; thus circulation regime is classified as a slow rotator. 
Although the condition of maintenance scenario against atmospheric collapse would be affected by the circulation regime, if planetary size is same as the Earth (as assumed in this study), circulation regime on tidally-locked planets would be slow rotator. 
Since TRAPPIST-1 we assumed have relatively small mass and low luminosity compared to other M dwarfs, most of other tidally-locked M dwarf planets to be observed would have slower rotation rate than TRAPPIST-1 planets. 

Unless planetary size is enough large to change the circulation regime from slow rotator to Rhines or fast rotator, atmospheric behavior such as surface temperature distribution and day-night heat transport is not be changed largely; thus, the maintenance scenario where surface liquid water remains against atmospheric collapse shown in this study would be found on planets around other M dwarfs.

\subsection{\ce{CO2} as a variable gas}

In our simulations, \ce{CO2} can condense in the atmosphere or on the surface,  but it is not treated as a variable gas species in terms of atmospheric composition. 
This is because the model we used can treat only one variable species (\ce{H2O} in our simulations). 
In other words, the ratio of \ce{N2} and \ce{CO2} in the dry air component remains constant everywhere, even though \ce{CO2} condensation locally alters the composition.
This limitation is why we did not perform continuous calculations to track $p_{\ce{CO2}}$ until atmospheric collapse ceased.
The fate of $p_{\ce{CO2}}$ during/after atmospheric collapse could be estimated by comparing each $p_{\ce{CO2}}$ case in Figure \ref{fig:collapse} if atmospheric \ce{CO2} decreases quasi-statically. 
On the other hand, if atmospheric collapse occurs rapidly, \ce{CO2} condensation can lead to an inhomogeneous \ce{CO2} mixing ratio. 
This inhomogeneity could result in a weaker \ce{CO2} greenhouse effect on the nightside compared to the dayside, for example.
Since lower nightside greenhouse effect results in lower minimum surface temperature, the $p_{\ce{CO2}}$ after atmospheric collapse is expected to be lower than the value we showed. 
Lower $p_{\ce{CO2}}$ after atmospheric collapse would make cooler dayside temperature. 

Furthermore, including multi-condensable species would be crucial for simulating cooler planets, especially those located farther from their star. This is because other species, such as \ce{CH4} and \ce{N2}, can also condense in such atmospheres.
To accurately simulate the atmosphere during atmospheric collapse, it would be needed to modify radiative transfer scheme and convection scheme. 
In future work, we develop new schemes that can account for these processes and investigate the detailed mechanisms of atmospheric collapse.

\subsection{Climatic evolution and onset of atmospheric collapse}

Changing atmospheric condition such as $p_{\ce{N2}}$ and $p_{\ce{CO2}}$ through climatic evolution (not only atmospheric collapse but also other components such as degassing and atmospheric escape) can be interpreted as orbits of the climate system in a phase space. 
According to the results in Figure \ref{fig:collapse}, the left-side boundary between the red region and blue region (atmospheric collapse occurs and doesn't occur) is a bifurcation point where atmospheric collapse begins. 
If a point in the phase space which has high $p_{\ce{CO2}}$ reaches the bifurcation point, the atmospheric composition would transit to the the right-side boundary between the red region and blue region, but $p_{\ce{CO2}}$ would hardly return to the \ce{CO2} rich atmosphere because of the one-way transition.

There are two cases where the one-way transition occurs: (1) a point reaches the bifurcation point, and (2) the bifurcation point changes. 
The first case is corresponding to the decreasing in $p_{\ce{CO2}}$ through climatic evolution by changing in such as degassing rate and weathering \citep{graham2022-01, graham2024-01}. 
The second case is the bifurcation point change where $p_{\ce{CO2}}$ remains but others change. 
For example, episodic volcanic event is possible to decrease surface temperature by reflecting stellar insolation \citep{macdonald2017-01}. 
Temporal temperature decrease induced by the episodic event would lead atmospheric collapse if the lowest surface temperature reached the \ce{CO2} condensation temperature. If this episodic event persists long enough, atmospheric collapse leads to lower $p_{\ce{CO2}}$ state than the original $p_{\ce{CO2}}$, resulting in a transition to other $p_{\ce{CO2}}$ state. Returning to the climate before the sudden episodic event would be inhibited. On the other hand, in case where the duration of temperature decrease is short, the surface temperature could return to the original state before the onset of atmospheric collapse. 
Evolution of stellar luminosity also changes the bifurcation point of the onset of atmospheric collapse. Different from the Sun-like stars, M-type stars' luminosity decreases after pre-main sequence \citep{baraffe2002-01}. 
In addition to the evolution of stellar luminosity corresponding to the planetary surface temperature, the evolution of M-type star could also have an influence to the onset of atmospheric collapse by changing the atmospheric composition. The evolution of M-type star also put photochemical constrains on the atmospheric composition \citep{gao2015-01}. 
Since the $p_{\ce{CO2}}$ dependency on the \ce{CO2} condensation temperature, long-term atmospheric evolution by photochemical reaction could cause atmospheric collapse.

\subsection{Outer edge of habitable zone for tidally-locked planets}

Although the outer edge of the classical HZ is defined by high \ce{CO2} greenhouse effect, we showed that tidally-locked planets can sustain locally habitable environment if atmospheric \ce{CO2} condenses. 
It suggests that in addition to high \ce{CO2} atmosphere, planets with low \ce{CO2} atmosphere can also be an additional type of habitable planets if the planet has small total mass of atmosphere and high day-night temperature contrast.

Besides, we didn't consider the effect of ocean. 
Exoplanets around M dwarfs are thought to have various amount of surface water \citep{tian2015-02, kimura2022-01}. 
Some of the planets would have larger ocean mass than the Earth, which type of exoplanets is known as ocean planets \citep{leger2004-01}. 
Ocean heat transport would decrease the day-night temperature contrast as well as large mass of background atmosphere (\ce{N2} in our simulations). 
However, the effect of ocean heat transport depends on such as disturbance by land distribution \citep{salazar2020-01, schmidt2022-01} and sea ice drift \citep{yang2020-01}. 
In addition to the heat transport, ocean as a \ce{CO2} storage makes the climatic behavior more complex. 
Additional processes of \ce{CO2} cycle (i.e. atmosphere-ocean \ce{CO2} exchange, burying the \ce{CO2} ice into the ocean, and formation of \ce{CO2} ice clathrate) should be considered.

\section{Summary} 

The classical condition of the outer edge of the habitable zone requires thick \ce{CO2} atmosphere to obtain strong greenhouse effect. 
However, in case of tidally-locked planets around M dwarfs, thick \ce{CO2} atmosphere would be condensed due to the low nightside temperature. 
This phenomenon known as atmospheric collapse could be an obstacle to the planetary habitability of cooler tidally-locked planets. 
On the other hand, surface temperature distribution depends on not only greenhouse effect but also global atmospheric heat transport from dayside to nightside. 
Although atmospheric collapse results in the loss of greenhouse effect, day-night atmospheric heat transport could be decrease largely due to the loss of total pressure during atmospheric collapse.  
This reduction would maintain local habitable environment on the dayside by leading to less heat redistribution of strong dayside insolation to the nightside.  

In this paper, we investigated the impact of atmospheric collapse to the planetary habitability on tidally-locked planets using a global climate model, the Generic PCM.
As a result, in case where the insolation is $0.4$ of the Solar constant ($S_{\mathrm{p}} = 0.4 \, S_{0}$) and $p_{\ce{CO2}} < 0.1$ \si{bar}, atmospheric \ce{CO2} condenses in the nightside high latitudes where surface temperature is the lowest. 
If \ce{CO2} condensation continues quasi-statically, $ p_{\ce{CO2}} $ after the atmospheric collapse event would be $ 10^{-3} \, \si{bar} < p_{\ce{CO2}} < 10^{-2} \, \si{bar} $ and $ 10^{-7} \, \si{bar} < p_{\ce{CO2}} < 10^{-6} \, \si{bar} $ in case with $p_{\ce{N2}} = 1.0$ \si{bar} and $p_{\ce{N2}} = 0.1$ \si{bar}, respectively. 

In both cases of $ p_{\ce{N2}} $, dayside surface liquid water could remain during/after the atmospheric collapse event against decreasing in \ce{CO2} greenhouse effect. 
We confirmed that atmospheric collapse also weakens the efficiency of day-night atmospheric heat transport due to loss of total atmospheric amount, and then, the reduction results in the maintenance of locally habitable environment. 

According to the traditional theory of the habitable zone (HZ), the definition of outer edge requires a massive \ce{CO2} atmosphere. 
Our results indicate that a high $p_{\ce{CO2}}$ atmosphere is unlikely to persist if atmospheric collapse occurs. 
However, a locally habitable environment on the dayside may be sustained despite a reduced \ce{CO2} greenhouse effect. 
Regarding cooler tidally-locked planets around the outer edge of HZ, we identify two potential types of habitable planets: those with a high $p_{\ce{CO2}}$ atmosphere, consistent with the classical habitable zone (HZ), and those with a low $p_{\ce{CO2}}$ atmosphere, resulting in a larger day-night temperature contrast. 
Our finding provides a revised understanding of the climate on tidally-locked planets. 
Furthermore, it also provides the necessity to consider three-dimensional structures of climate such as heat distribution and global atmospheric circulation.

\begin{acknowledgments}
This work was supported by a Grant-in-Aid for JSPS Fellows (No. 23KJ0940) from the Japan Society for the Promotion of Science (JSPS). 
T. Kodama acknowledges Strategic Research Projects grant from Research Organization of Information and Systems (ROIS) (No.2024-SRP-3), the Astrobiology Center of National Institutes of Natural Sciences (NINS) (No. AB0608, AB0708), and a grant-in-aid of research from Itoh Science Foundation. 
T. Kodama also acknowledges a Grant-in-Aid for Challenging Research (Exploratory) (No. 23K17709) from the JSPS. 
M. Turbet acknowledges support from the Tremplin 2022 program of the Faculty of Science and Engineering of Sorbonne University. 
M. Turbet also acknowledges support from BELSPO BRAIN (B2/212/PI/PORTAL).
G. Chaverot acknowledges the financial support of the SNSF (grant number: P500PT\_217840). 
K. Taniguchi, T. Kodama, and H. Genda acknowledge a Fund for the Promotion of Joint International Research (International Leading Research) (No. 22K21344) from the JSPS. 
This study was also supported by the Cooperative Program (JURCAOSCFG24-06, JURCAOSCFG25-16) of Atmosphere and Ocean Research Institute, The University of Tokyo and the PPARC joint research program of Tohoku University.

\end{acknowledgments}

\bibliography{reference}{}
\bibliographystyle{aasjournalv7}

\end{document}